\documentclass[lettersize,journal]{IEEEtran}
\usepackage{graphicx}
\usepackage{graphics}
\usepackage{colortbl}
\usepackage{multicol}
\usepackage{diagbox}
\usepackage{lipsum}
\usepackage{color}
\usepackage[cmex10]{amsmath}
\usepackage{amsthm}
\usepackage{amssymb}
\usepackage{mathrsfs}
\usepackage{mathtools}
\usepackage{amsbsy}
\usepackage[colorlinks=true,bookmarks=false,citecolor=blue,urlcolor=blue]{hyperref}
\usepackage[dvipsnames]{xcolor}\usepackage{mathrsfs}
\usepackage{setspace}
\usepackage{float}
\usepackage{graphicx}
\usepackage{pstool}
\usepackage{epstopdf}
\usepackage{lettrine}
\usepackage{psfrag}
\usepackage{newclude}
\usepackage[normalem]{ulem}
\usepackage{latexsym}
\usepackage{algpseudocode}
\usepackage{algorithm,algpseudocode}
\usepackage{algorithmicx}
\usepackage{gensymb}
\usepackage{marginnote}
\usepackage{multirow}
\usepackage{amsmath}
\usepackage{amsmath,bm}
\usepackage{wasysym}
\usepackage{mathrsfs}
\usepackage{amsmath,cite,amsfonts,amssymb,color}
\usepackage[T1]{fontenc}
\usepackage{graphicx,psfrag}
\usepackage{pstool}
\usepackage{algorithm}
\usepackage{algpseudocode}
\usepackage{comment}
\usepackage{tabularx}
\usepackage{caption}
\usepackage{csquotes}
\usepackage{float}
\usepackage{lettrine}
\usepackage{makecell}
\usepackage{textcomp, gensymb}
\usepackage{relsize}
\usepackage{balance}

\ifCLASSOPTIONcompsoc
\usepackage[caption=false,font=normalsize,labelfon
t=sf,textfont=sf]{subfig}
\else
\usepackage[caption=false,font=footnotesize]{subfig}
\fi

\allowdisplaybreaks

\graphicspath{{figures/}}
\allowdisplaybreaks

\newcommand{\RNum}[1]{\uppercase\expandafter{\romannumeral #1\relax}}

\usepackage{mathtools}

\usepackage{comment}
\begin{document}

\title{On Age of Information and Energy-Transfer in a STAR-RIS-assisted System}

\author{\IEEEauthorblockN{Mohammad Reza Kavianinia, Mohammad Mehdi Setoode, and Mohammad Javad Emadi\thanks{Mohammad Reza Kavianinia, Mohammad Mehdi Setoode and Mohammad Javad Emadi are with the Department of Electrical Engineering, Amirkabir University of Technology (Tehran Polytechnic), Tehran, Iran (E-mails: \{mrezakaviani, mohammad.setoode, mj.emadi\}@aut.ac.ir).}
}}



\maketitle

\begin{abstract}
Battery-limited devices and time-sensitive applications are considered as key players in the forthcoming wireless sensor network. Therefore, the main goal of the network is two-fold; Charge battery-limited devices, and provide status updates to users where information-freshness matters. In this paper, a multi-antenna base station (BS) in assistance of simultaneously-transmitting-and-reflecting reconfigurable intelligent surface (STAR-RIS) transmits power to energy-harvesting devices while controlling status update performance at information-users by analyzing age of information (AoI) metric. Therefore, we derive a scheduling policy at BS, and analyze joint transmit beamforming and amplitude-phase optimization at BS and STAR-RIS, respectively, to reduce average sum-AoI for the time-sensitive information-users while satisfying minimum required energy at energy-harvesting users. Moreover, two different energy-splitting and mode-switching policies at STAR-RIS are studied. Then, by use of an alternating optimization algorithm, the optimization problem is studied and non-convexity of the problem is tackled by using the successive convex approximation technique. Through numerical results, AoI-metric and energy harvesting requirements of the network are analyzed versus different parameters such as number of antennas at BS, size of STAR-RIS, and transmitted power to highlight how we can improve two-fold performance of the system by utilizing STAR-RIS compared to the conventional RIS structure.
\end{abstract}


\vspace{-0.0 cm}
\section{Introduction}
\lettrine{E}{nergy} harvesting is a highly appealing technology for a variety of self-powered devices that has gained significant interest. Internet of things (IoT) is one example of such systems which are entangled with the energy harvesting feature. However, if energy harvesting procedure takes a long period, the user's performance/experience may not be acceptable due to the  delays\cite{hu2020aoi}. Moreover, some other emerging applications, such as cooperative autonomous driving also require fresh and real-time status information \cite{sorkhoh2021optimizing}. Therefore, energy harvesting and data freshness are two important features in the forthcoming wireless networks.

\begin{figure}[t!]
    \centering
    \pstool[scale=0.42]{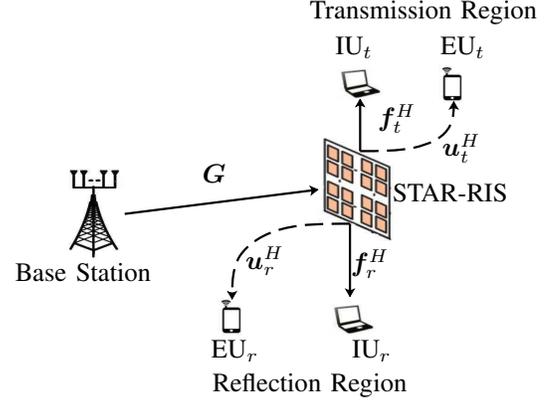}{
    \psfrag{A}{\hspace{-1cm}$\text{Base Station}$}
    \psfrag{B}{\hspace{-0.35cm} $\text{EU}_r$}
    \psfrag{C}{\hspace{-0.05cm}$\text{IU}_r$}
    \psfrag{D}{\hspace{-0.3cm} $\text{IU}_t$}
    \psfrag{E}{\hspace{-0.25cm} $\text{EU}_t$}
    \psfrag{F}{\hspace{-0.45cm} \text{STAR-RIS}}
    \psfrag{G}{\hspace{-0.4cm}$\boldsymbol{G}$}
    \psfrag{H}{\hspace{-0.3cm} $\boldsymbol{u}_r^H$}
    \psfrag{I}{\hspace{-0.3cm}$\boldsymbol{f}_r^H$}
    \psfrag{J}{\hspace{-0.1cm}$\boldsymbol{f}_t^H$}
    \psfrag{K}{\hspace{-0.15cm}$\boldsymbol{u}_t^H$}
    \psfrag{L}{\hspace{-1cm}$\text{Reflection Region}$}
    \psfrag{M}{\hspace{-1.2cm}$\text{Transmission Region}$}
    }
    \caption{STAR-RIS-assisted downlink SWIPT system.}
    \label{fig:1}
    \vspace{-.65cm}
\end{figure}

One of the key technology to improve
performance of a wireless system, e.g. spectral-efficiency, energy-efficiency and coverage, is the reconfigurable intelligent surface (RIS). By controlling the phase \textcolor{black}{and/or amplitude} of each elements of RIS, one can control the wavefront or equivalently manipulate the wireless propagation channel to enhance the performance \cite{basar2019wireless}. Recently, a promising version of RIS, known as simultaneously transmitting and reflecting RIS (STAR-RIS) has evolved, wherein each element of the RIS can partly reflect and/or transmit the waveform while individually controlling amplitudes and phases of transmitted and reflected parts. Compared to the conventional RIS, one of the main advantages of STAR-RIS is enhancing the coverage from half-space to full-space \cite{liu2021star}. Numerous studies on STAR-RIS assisted wireless networks are  published in the literature. In \cite{du2023star}, a wireless powered IoT network based on a STAR-RIS is studied which indicates the STAR-RIS enhances energy harvesting and information transfer for the considered setup. Also, a wireless-powered mobile edge computing system aided STAR-RIS is investigated in \cite{qin2023joint}, where an STAR-RIS is deployed to assist the energy transfer in the downlink as well as the task offloading in the uplink. Recently, STAR-RIS assisted simultaneous wireless information and power transfer (SWIPT) system is studied in \cite{liu2022effective}. The trade-off between information users and energy users is investigated through a multi-objective optimization problem.

On the other hand, age of information (AoI), a new metric to measure freshness of information, is defined as the amount of time elapsed from the generation of a previous successfully transmitted status update packet \cite{abd2020aoi}. AoI is becoming one of the key metrics in telecommunication systems which has been investigated and analyzed for different systems and scenarios \cite{yates2021age}. For instance, average peak of AoI for a two-hop relay wireless power transfer based system is studied in \cite{khorsandmanesh2021average}, wherein a source and a relay capture their required power by harvesting from  received signal transmitted from a power station. Moreover, \cite{lyu2022optimizing} has investigated potential impact of the conventional RIS in SWIPT system on minimizing sum-AoI. 

Most of available studies on STAR-RIS mainly have focused on achievable rate and energy efficiency improvements of a system, and information freshness analysis in presence of STAR-RIS aided SWIPT networks is not studied in the literature. Therefore, in this paper, we present a STAR-RIS assisted SWIPT system in presence of two types of users: 1- Information users (IUs) that require fresh information 2- Energy users (EUs) which ask for energy harvesting from a base station. Also, thanks to the STAR-RIS and 360 degree coverage, in contrast to the conventional RIS technique, one can support users in front and behind of the STAR-RIS, as well. Thus, we study minimizing of average sum-AoI subject to energy harvesting constraints for different STAR protocols. A non-convex problem is formulated to optimize transmit beamforming, phase shifts, and the proposed scheduling scheme. The problem is solved by a successive convex approximation (SCA) based alternating optimization (AO) algorithm. Moreover, numerical results for different protocols of STAR-RIS are evaluated and indicate performance improvements compared to that of the conventional RIS.

\indent\textit{Organization}: Section II represents the proposed system model, STAR-RIS configuration and defines AoI formulation. Section III introduces the optimization problem and specifies the proposed algorithm. Section IV presents the numerical results, and  section V concludes the paper.

\indent\textit{Notations}: $\text{tr}(\boldsymbol{F})$ denotes trace of matrix $\boldsymbol{F}$, $\boldsymbol{F} \succeq 0$ declares that $\boldsymbol{F}$ is a positive semi-definite matrix, and an element in $i$-th row and $j$-th column is represented by $\boldsymbol{F}[i,j]$.
$\big|\boldsymbol{x}\big|$ specifies the Euclidean norm of the complex-valued vector $\boldsymbol{x}$, and real part of $x$ is $\mathfrak{R}\{x\}$. Probability of a random variable $x$ is denoted by Pr$\{x\}$, and a zero-mean and unit-variance complex symmetric Gaussian random variable is $x \sim \mathcal{C}\mathcal{N} (0,1)$.

\section{System Model}
A downlink STAR-RIS-aided SWIPT wireless system depicted in Fig. \ref{fig:1}, in which a base station (BS) with $N_t$ antennas, by utilizing the STAR-RIS, transmits signals towards single-antenna energy users and information users. The STAR-RIS is equipped with $M$ elements and is utilized to improve performance and coverage of the system by tuning transmitting and reflecting coefficients (TaRCs). As depicted in the figure and without loss of generality and for simplicity, it is assumed that two energy users, i.e. $\text{EU}_t$ and $\text{EU}_r$, and two information users $\text{IU}_t$ and $\text{IU}_r$ are located in the transmission and reflection regions of STAR-RIS, respectively. 
 
Also, we consider one information stream per information user and one energy stream for each energy user. Therefore, the BS sends energy signals and status update packets to the energy users and the information users over energy and information channels, respectively. It is assumed that energy channels are orthogonal to information channels to control interference over the information channel and enhance signal-to-noise ratio (SNR). Moreover, it is assumed that there are $N$ time slots 
 indexed by $n \in \{1,…,N\}$, and there is one information channel per time slot. Thus, the scheduling policy to support information users at the BS is constrained by
\begin{eqnarray}
      &&  s_{t}(n) + s_{r}(n) \leq 1, \label{eq:1} \\
 && s_{k}(n) \in \{0,1\}, \forall  k \in \{t, r\}, \label{eq:2}
\end{eqnarray}
where $s_{k}(n)$ denotes scheduling state at time slot $n$ for $\text{IU}_k$.
\raggedbottom
\subsection{STAR-RIS Configuration}
We consider the two well-known energy splitting (ES) and mode switching (MS) schemes for the STAR-RIS \cite{liu2021star}.
In the ES protocol, all STAR-RIS elements are able to operate in transmitting and reflecting modes at the same time by splitting the energy. In contrast, for the MS case, each element can operate only in transmitting or reflecting mode. The TaRCs of a generic STAR-RIS are denoted by
$\boldsymbol{\Phi}_k^{\text{ES/MS}}=\textrm{diag}(\sqrt{\alpha_1^k}e^{j\theta_1^k}, \dots , \sqrt{\alpha_M^k}e^{j\theta_M^k})$, $\forall k \in \{t, r\}$, where $\alpha_m^t,\alpha_m^r \in \left[0,1\right]$ for ES and $\alpha_m^t,\alpha_m^r \in \{0,1\}$ for MS, and $\theta_m^t,\theta_m^r \in \left[0,2\pi\right), \forall m\in M$; These parameters represent amplitude and phase
shift of the $m$-th element, respectively. It is worth noting that the amplitude parts of TaRCs are subject to law of energy conservation, i.e. $\alpha_m^t + \alpha_m^r =1$ \cite{liu2021star,xu2021star}.

\subsection{Signal Transmissions and Receptions}
BS sends $\boldsymbol{t}_k^I(n)\!\!=\!\!\boldsymbol{\omega}_k(n)x_k^I(n)$ and $\boldsymbol{r}_k^E(n)\! =\!\boldsymbol{\vartheta}_k(n)x_k^E(n)$ to $\text{IU}_k$ and $\text{EU}_k$, wherein  $\boldsymbol{\omega}_k(n) \in \mathbb{C}^{N_t \times 1}$ and $\boldsymbol{\vartheta}_k(n) \in \mathbb{C}^{N_t \times 1}$ denote beamforming vectors for $\text{IU}_k$ and $\text{EU}_k$, respectively. Also,  information and energy signals are presented by $x_k^I(n)\sim \mathcal{C}\mathcal{N} (0,1)$ and $x_k^E(n) \sim \mathcal{C}\mathcal{N} (0,1)$, respectively. By assuming  maximum power constraint $P_o$ at the BS, we have 
\begin{equation}
    \sum_{k \in \{t, r\}} {\big| \boldsymbol{\omega}_k(n) \big|}^2 + \sum_{k \in \{t, r\}} {\big| \boldsymbol{\vartheta}_k(n) \big|}^2  \le P_o.
    \label{eq:3}
\end{equation}

Therefore, the single-antenna $k$-th information user receives
 \begin{equation}
    y_{k}^{\text{IU}}(n)= \boldsymbol{f}_k^H(n)\boldsymbol{\Phi}_k^{\text{ES/MS}}(n)\boldsymbol{G}(n)\boldsymbol{t}_k^I(n) + n_{k}^{\text{IU}}, 
    \end{equation}
where $\boldsymbol{G}(n) \in \mathbb{C}^{M \times N_t}$ denotes the channel between the BS and STAR-RIS, $\boldsymbol{f}_k^H(n) \in \mathbb{C}^{1 \times M}$ is the channel between STAR-RIS and $\text{IU}_k$, $\boldsymbol{\Phi}_k^{\text{ES/MS}}(n)\in \mathbb{C}^{M \times M}$ indicates STAR-RIS settings, and $n_{k}^{\text{IU}}\sim \mathcal{C}\mathcal{N} (0,\sigma_{I,k}^2)$ denotes the additive white Gaussian noise (AWGN) at the information user $\text{IU}_k$. Thus, the SNR at $\text{IU}_k$ becomes
\begin{equation}
\gamma_k(n)=\frac{{\big|\boldsymbol{f}_k^H(n)\boldsymbol{\Phi}_k^{\text{ES/MS}}(n)\boldsymbol{G}(n)\boldsymbol{\omega}_k(n) \big|}^2}{ \sigma_{I,k}^2}.
\label{eq:5a}
\end{equation}

Similarly, the $k$th single-antenna energy user receives
\begin{equation}
    y_{k}^{\text{EU}}(n)= \boldsymbol{u}_k^H(n)\boldsymbol{\Phi}_k^{\text{ES/MS}}(n)\boldsymbol{G}(n)\boldsymbol{r}_k^E(n) + n_{k}^{\text{EU}}, 
    \end{equation}
where $\boldsymbol{u}_k^H(n) \in \mathbb{C}^{1 \times M}$ is the channel between STAR-RIS and $\text{EU}_k$, and $n_{k}^{\text{EU}}\sim \mathcal{C}\mathcal{N} (0,\sigma_{E,k}^2)$ indicates AWGN at the energy user $\text{EU}_k$. By neglecting noise power relative to the received signal, the harvested energy at $\text{EU}_k$ becomes
\begin{equation}
E_k(n)={\big|\boldsymbol{u}_k^H(n)\boldsymbol{\Phi}_k^{\text{ES/MS}}(n)\boldsymbol{G}(n)\boldsymbol{\vartheta}_k(n) \big|}^2.
\label{eq:7a}
\end{equation}



\subsection{Age of Information}
The AoI metric indicates the freshness of data of a source from the destination's perspective. $A_{k}(n)$ denotes the AoI of $\text{IU}_{k}$ in $n$-th time slot for all $n \in \{ 1,...,N \}$ and $k \in \{ t , r \}$, which is given by \cite{muhammad2023age}
\begin{equation}
\begin{split}
    A_{k}(n+1) = s_{k}(n)b_{k}(n)z_{k}(n)+(1-s_{k}(n)b_{k}(n))A_{k}(n) + 1, \label{eq:4}
\end{split}
\end{equation}
where $z_{k}(n)$ is the system time of the packet in the stream corresponding to $\text{IU}_{k}$, named as $\text{Str}_{k}$. That is, 
\vfill
\begin{equation}
    z_{k}(n+1) = \left\{\begin{matrix}
    0 &   p_{k}(n+1)=1 \\ 
    z_{k}(n) + 1 & \text{o.w.}
    \end{matrix}\right. \label{eq:5}
\end{equation}
\textcolor{black}{where the binary random variable $p_{k}(n)$ is equal to 1 when a \textit{new} packet from $\text{Str}_{k}$ arrives at the BS at time slot n, with a probability of $\lambda_{k}$; $\text{Pr} \{ p_{k}(n)=1 \} = \lambda_{k}$. Also, $b_{k}(n) \in \{ 0,1 \}$ is 1 when $\text{Str}_{k}$ has an available packet to transmit and turns to 0 only if $\text{Str}_{k}$ is scheduled and its packet is delivered successfully without any newly arrived packet.} So, $b_{k}(n)$ can be rewritten as follows
\begin{equation}
    b_{k}(n+1)=p_{k}(n+1)+b_{k}(n)(1-s_{k}(n))(1-p_{k}(n+1)). \label{eq:6}
\end{equation}

Moreover, a packet from $ \text{Str}_{k}$ can only be successfully transmitted when it is scheduled and the received SNR to $\text{IU}_k$ is greater than the threshold value, $\gamma_{th}$ as follows
\begin{equation}
    \gamma_{k}(n) \geq s_{k}(n)b_{k}(n)\gamma_{th} , \forall  k \in \{t, r\}. \label{eq:7}
\end{equation}

\section{Problem Formulation and Optimization}
We minimize the average sum-AoI for all streams over the time horizon to satisfy the freshness for information users, subject to transferring enough energy to the energy harvesting users. Hence, the optimization problem for ES protocol can be formulated as
\begin{subequations}
\label{P1}
\begin{align} 
\mathcal{P}_{1}:\min_{\substack{s_k(n), \boldsymbol{\Phi}_k(n) \\  \boldsymbol{\omega}_k(n),\boldsymbol{\vartheta}_k(n)}}
\quad & \frac{1}{2N} \sum_{n}\sum_{\forall k \in \{ t,r \}} A_{k}(n)\\
\textrm{s.t.}
\quad & \eqref{eq:1}, \eqref{eq:2}, \eqref{eq:3}, \eqref{eq:7}, \label{eq:8b} \\
\quad & \big[\boldsymbol{\Phi} \big]_m = \sqrt{\alpha_m^k}e^{j\theta_m^k}, \forall  m \in M,  \label{eq:8c} \\
\quad & \alpha_m^k \in \big[0,1 \big], \forall  k \in \big\{t,r\big\},\label{eq:8d} \\
\quad & \alpha_m^t + \alpha_m^r = 1,\forall  m \in M, \label{eq:8f}\\
\quad & \theta_m^t,\theta_m^r \in \left[0,2\pi\right), \forall m\in M, \label{eq:8g}\\
\quad & E_k(n) \ge E, \forall  k \in \big\{t,r\big\}, \label{eq:8h}  
\end{align}
\end{subequations}
where the constraints \eqref{eq:8c}-\eqref{eq:8g} denote STAR-RIS protocols configurations and the constraint \eqref{eq:8h} indicates the required minimum harvested energy $E$. 

\textcolor{black}{By definition, the AoI increases linearly until a status update packet is successfully received, and according to \cite{zhang2020age}, the problem of minimizing the average sum-AoI is equivalent to maximizing the AoI reduction in each time slot, i.e.  the difference between the AoI and the system time in that particular slot \cite{lyu2022optimizing}, and the optimization problem \eqref{P1} is reformulated as follows}
\begin{subequations}
\label{P2}
\begin{align}
\max_{\substack{s_k(n), \boldsymbol{\Phi}_k(n) \\  \boldsymbol{\omega}_k(n),\boldsymbol{\vartheta}_k(n)}} \quad & \sum_{\forall k \in \{ t,r \}} (A_k(n)-z_k(n))s_k(n)b_k(n)\\
\textrm{s.t.}
\quad & \eqref{eq:8b}-\eqref{eq:8h}.
\end{align}
\end{subequations}

Because of the non-convex constraints, i.e. \eqref{eq:2},~\eqref{eq:7},~\eqref{eq:8h} and STAR-RIS settings, the optimization problem \eqref{P2} cannot be solved directly. To deal with the non-convexity issue, we reformulate \eqref{P2}, and an SCA algorithm based on AO approach is proposed and analyzed in what follows to solve the problem.

\subsection{Problem Reformulation}
To deal with the non-convexity of the quality of service and energy harvesting constraints, i.e. \eqref{eq:7} and \eqref{eq:8h} respectively, the equations \eqref{eq:5a} and \eqref{eq:7a} are reformulated as follows
\begin{equation}
\begin{aligned}
    &\boldsymbol{P}^k(n) = \text{diag}(\boldsymbol{f}_{k}^H(n))\boldsymbol{G}(n)\boldsymbol{\omega}_{k}(n),\\
    &\boldsymbol{V}^k(n) = \text{diag}(\boldsymbol{u}_{k}^H(n))\boldsymbol{G}(n)\boldsymbol{\vartheta}_{k}(n),
\end{aligned}
\end{equation}
and the TaRCs are rewritten as $\boldsymbol{l}^{t/r}=\big[\sqrt{\alpha_1^{t/r}}e^{j\theta_1^{t/r}}, \dots , \sqrt{\alpha_M^{t/r}}e^{j\theta_M^{t/r}}\big]^H$. Thus, the non-convex constraints, i.e. \eqref{eq:7} and \eqref{eq:8h}, are  reformulated as 
\begin{equation}
\begin{split}
    &\gamma_{k}(n)=\text{tr}(\boldsymbol{P}_I^k(n)\boldsymbol{\varphi}^k(n)) \geq s_{k}(n)b_k(n)\gamma_{th}, \forall  k \in \{t, r\},\\
    &E_{k}(n)=\text{tr}(\boldsymbol{V}_E^k(n)\boldsymbol{\varphi}^k(n)) \ge E, \forall  k \in \{t, r\}, 
    \label{eq:11} \raisetag{13pt}
\end{split}
\end{equation}
where $\boldsymbol{P}_I^k(n)=\boldsymbol{P}^k(n)(\boldsymbol{P}^k(n))^H$, $\boldsymbol{V}_E^k(n)=\boldsymbol{V}^k(n)(\boldsymbol{V}^k(n))^H$ and $\boldsymbol{\varphi}^k(n)=\boldsymbol{l}^k(\boldsymbol{l}^k)^H$, which are positive semi-definite matrices. Furthermore, the aforementioned reformulation problem imposes unit-rank matrice for  $\boldsymbol{\varphi}^k(n)$ which is a non-convex constraint. By using the semi-definite relaxation for removing the unit-rank constraints and relaxing \eqref{eq:2} as $s_{k}(n) \in [0,1]$, the problem \eqref{P2} is rewritten as
\begin{subequations}
\label{P3}
\begin{align}
\max_{\substack{s_{k}(n),\boldsymbol{\varphi}^k(n),\\  \boldsymbol{\omega}_{k}(n),\boldsymbol{\vartheta}_{k}(n),\\ \boldsymbol{\alpha}^t,\boldsymbol{\alpha}^r}} \quad & \sum_{\forall k \in \{ t,r \}} (A_{k}(n)-z_{k}(n))s_{k}(n)b_{k}(n)\\
\textrm{s.t.}
\quad & \eqref{eq:1},\eqref{eq:3},\eqref{eq:8d}-\eqref{eq:8g},\eqref{eq:11}, \\
\quad & s_{k}(n) \in [0,1], \forall k \in \{r, t\}, \\
\quad & \boldsymbol{\varphi}^k(n)[i,i]=\boldsymbol{\alpha}^k[i], \forall k \in \{r, t\},  \label{eq:12d}\\
\quad & \boldsymbol{\varphi}^k(n) \succeq 0, \forall k \in \{r, t\},\label{eq:12e}
\end{align}
\end{subequations} which is not a convex optimization problem with respect to the beamforming vectors and TaRCs.
Using the similar approach as presented in \cite{saeidi2021weighted,niu2021simultaneous}, the AO approach is provided to solve the optimization problem \eqref{P3} by splitting it into two sub-problems as discussed in what follows. 

\subsection{The Scheduling Policy and TaRCs Optimization for ES}
Due to the highly-coupled optimization problem \eqref{P3}, we split the main problem into two sub-problems. Firstly,  for given $\boldsymbol{\omega}_{k}(n)$ and $\boldsymbol{\vartheta}_{k}(n)$, we find out the scheduling indicator and TaRCs by solving the following problem as follows
\begin{subequations}
\label{P4}
\begin{align}
\max_{\substack{s_{k}(n),\boldsymbol{\varphi}^k(n),\\ \boldsymbol{\alpha}^t,\boldsymbol{\alpha}^r}} \quad & \sum_{\forall k \in \{ t,r \}} (A_{k}(n)-z_{k}(n))s_{k}(n)b_{k}(n)\\
\textrm{s.t.}
\quad & \eqref{eq:1},\eqref{eq:3},\eqref{eq:8d}-\eqref{eq:8g},\eqref{eq:11}, \\
\quad & s_{k}(n) \in [0,1],  \forall k \in \{r, t\},\label{eq:13c} \\
\quad & \boldsymbol{\varphi}^k(n)[i,i]=\boldsymbol{\alpha}^k[i], \forall k \in \{r, t\},  \label{eq:13d}\\
\quad & \boldsymbol{\varphi}^k(n) \succeq 0, \forall k \in \{r, t\}, \label{eq:13e}
\end{align}
\end{subequations}
which is a convex optimization problem.
\subsection{The Scheduling Policy and TaRCs Optimization for MS}
The only difference between ES and MS protocols is the constraint \eqref{eq:8d}, which is substituted by binary constraint and provides non-convex constraint. To deal with the binary constraint, an upper bound is calculated in the $j$-th iteration by the following first-order Taylor expansion \cite{mu2021simultaneously} 
\begin{equation}
    \begin{split}
         \alpha_m^k-(\alpha_m^k)^2 &\le (\alpha_m^{k(j)})^2 + (1-2\alpha_m^{k(j)})\alpha_m^k\\
        & = g(\alpha_m^{k(j)},\alpha_m^k), \forall m \in M, 
    \end{split}
    \label{eq:18}
\end{equation}
which the penalty term $-\mu\sum_{m=1}^{M}\sum_{k\in \{r, t\}}g(\alpha_m^{k(j)},\alpha_m^k)$ with large positive constant $\mu$, added to the objective function of \eqref{P4} to address non convexity of binary constraint.

\subsection{The Scheduling Policy and Beamforming Optimization with SCA Algorithm}
For the given optimal $\boldsymbol{\varphi}^k(n)$, the constraints \eqref{eq:11} are approximated by use of the Taylor expansion \cite{lyu2022optimizing} with respect to $\boldsymbol{\omega}_{k}(n)$ and $\boldsymbol{\vartheta}_{k}(n)$ as
\begin{equation}
 \begin{split}
     &2\mathfrak{R}\Big\{\boldsymbol{\omega}_{k}^{(i-1)H}(n)\text{tr}(\boldsymbol{p}_I^k(n)\boldsymbol{\varphi}^k(n))(\boldsymbol{\omega}_{k}(n)-\boldsymbol{\omega}_{k}^{(i-1)H}(n))\Big\}\\
     &+\text{tr}(\boldsymbol{p}_I^k(n)\boldsymbol{\varphi}^k(n))\big|\boldsymbol{\omega}_{k}^{(i-1)}(n)\big|^2 \geq s_{k}(n)b_k(n)\gamma_{th},\label{eq:15}
\end{split}
\raisetag{13pt}
\end{equation}

\begin{equation}
 \begin{split}
     &2\mathfrak{R}\Big\{\boldsymbol{\vartheta}_{k}^{(i-1)H}(n)\text{tr}(\boldsymbol{v}_E^k(n)\boldsymbol{\varphi}^k(n))(\boldsymbol{\vartheta}_{k}(n)-\boldsymbol{\vartheta}_{k}^{(i-1)H}(n))\Big\}\\
     &+\text{tr}(\boldsymbol{v}_E^k(n)\boldsymbol{\varphi}^k(n))\big|\boldsymbol{\vartheta}_{k}^{(i-1)}(n)\big|^2 \geq E,\label{eq:16}
\end{split}
\raisetag{13pt}
\end{equation}
where $\boldsymbol{p}_I^k(n)\!=\!\text{diag}(\boldsymbol{f}_{k}^H(n))\boldsymbol{G}(n)(\text{diag}(\boldsymbol{f}_{k}^H(n))\boldsymbol{G}(n))^H$, $\boldsymbol{v}_E^k(n) = \text{diag}(\boldsymbol{u}_{k}^H(n))\boldsymbol{G}(n)(\text{diag}(\boldsymbol{u}_{k}^H(n))\boldsymbol{G}(n))^H$ in the $(i-1)^\text{th}$ iteration. Thus, the optimization problem \eqref{P3} becomes
\begin{subequations}
\label{P5}
\begin{align}
\max_{\substack{s_{k}(n),\\ \boldsymbol{\omega}_{k}(n),\boldsymbol{\vartheta}_{k}(n)}} \quad & \sum_{\forall k \in \{ t,r \}} (A_{k}(n)-z_{k}(n))s_{k}(n)b_{k}(n)\\
\textrm{s.t.}
\quad & \eqref{eq:1},\eqref{eq:3},\eqref{eq:13c},\eqref{eq:15}, \eqref{eq:16},
\end{align}
\end{subequations}
which is a convex problem.
Finally, by utilizing AO between \eqref{P4} and \eqref{P5} for the ES protocol and generalized \eqref{P4} and \eqref{P5} for the MS one,  solutions are achieved as the algorithm converges. The iterative strategy for solving the problem \eqref{P3} is summarized in \textbf{Algorithm 1}.
\subsection{Complexity Analysis}
In the previous subsection, the optimization problems \eqref{P4} and \eqref{P5} for the ES and MS protocols are formulated, and can be  solved by CVX tool \cite{grant2014cvx}. The complexity of the proposed scheme for the ES protocol is $\mathcal{O}\big(T_{AO}(T_{S}M^{3.5}\log_2(1/\epsilon)))$, where $T_{AO}$, $T_{S}$ and $\epsilon$ are the number of alternating iterations, the number of SCA iterations and stopping accuracy of SCA scheme, respectively. Also, for the MS protocol, we have $\mathcal{O}\big(T_{AO}(I_BT_{S}M^{3.5}\log_2(1/\epsilon)))$, where $I_B$ denotes the number of iterations for the binary convergence. 
\begin{algorithm}[t!]
\caption{SCA  method for addressing optimization problem given in \eqref{P3}}\label{alg:cap2}
\textbf{Input}: Initial values for $\boldsymbol{\omega}_{k}^{(0)}(n)$ and $\boldsymbol{\vartheta}_{k}^{(0)}(n)$, Channel coefficients $\boldsymbol{G}$, $\boldsymbol{u}_{k}^H$, and $\boldsymbol{f}_{k}^H$. Maximum power $P_{o}$. Initial value for tolerance $\epsilon = 10^{-3}$. 
\begin{algorithmic}[1]
\For{$i=1, 2, \dots$}
    \State
    \parbox[t]{\dimexpr\linewidth-\algorithmicindent}{%
    For given ${\boldsymbol{\omega}_{k}}^{(i)}(n), \boldsymbol{\vartheta}_{k}^{(i)}(n),\forall k$ update $\boldsymbol{\varphi}^k(n)$ and $s_k(n)$ by solving $\mathcal{P}_{2.1}$.
    }
    \For{$j=1, 2, \dots$}
    \State
    \parbox[t]{\dimexpr\linewidth-\algorithmicindent}{%
    For given $\boldsymbol{\varphi}^k(n)$ and $s_k(n)$, solve the problem\\ $\mathcal{P}_{2.2}$ to update value of ${\boldsymbol{\omega}_{k}}^{(i + 1)}(n)$ and $\boldsymbol{\vartheta}_{k}^{(i + 1)}(n)$ 
    }
    \State
    \parbox[t]{\dimexpr\linewidth-\algorithmicindent}{%
    \textbf{Until}: Convergence of ${\boldsymbol{\omega}_{k}}(n)$, $\boldsymbol{\vartheta}_{k}(n)$ and $s_k(n)$ 
    }
    \EndFor
    \parbox[t]{\dimexpr\linewidth-\algorithmicindent}{%
    }
    \State \textbf{Until} Convergence of $\boldsymbol{\varphi}^k(n)$ and $s_k(n)$ .
\EndFor
\end{algorithmic}
\textbf{Output}: The optimal values: $\boldsymbol{\varphi}^k(n)$, $s_k(n)$, ${\boldsymbol{\omega}_{k}}(n)$ and $\boldsymbol{\vartheta}_{k}(n)$.
\end{algorithm}

\section{Numerical Results and Discussion}
To present numerical results, it is assumed that the  Cartesian coordinates of STAR-RIS, $\text{EU}_{r}$, $\text{IU}_{r}$, $\text{EU}_{t}$, $\text{IU}_{t}$ and  BS are $(8,0), (10,-2),  (12,-2),  (10,2),  (12,2)$  and $(0,0)$, respectively. Probabilities of the packet arrivals are $\lambda_k = 0.6$, the number of time slots is $N=100$, and the noise variance is $\sigma_{I,k}^2 = 1$. The path loss exponents for information and energy users are $\alpha_i = 2.2$ and $\alpha_e =2$, respectively. We also assume a stronger channel for the energy users than that of the information users. The geometric path-loss and random phase are considered for channel simulation as $\boldsymbol{F}=\sqrt{(1/{d})^\alpha}\boldsymbol{q}$,  where $d$ represents a relative distance and $\boldsymbol{q}=e^{j\phi}$ denotes a random phase shift with a uniform distribution over $\big[0, 2\pi \big]$.

\begin{figure}[t!]
    \centering
    \includegraphics[width=0.45\textwidth]{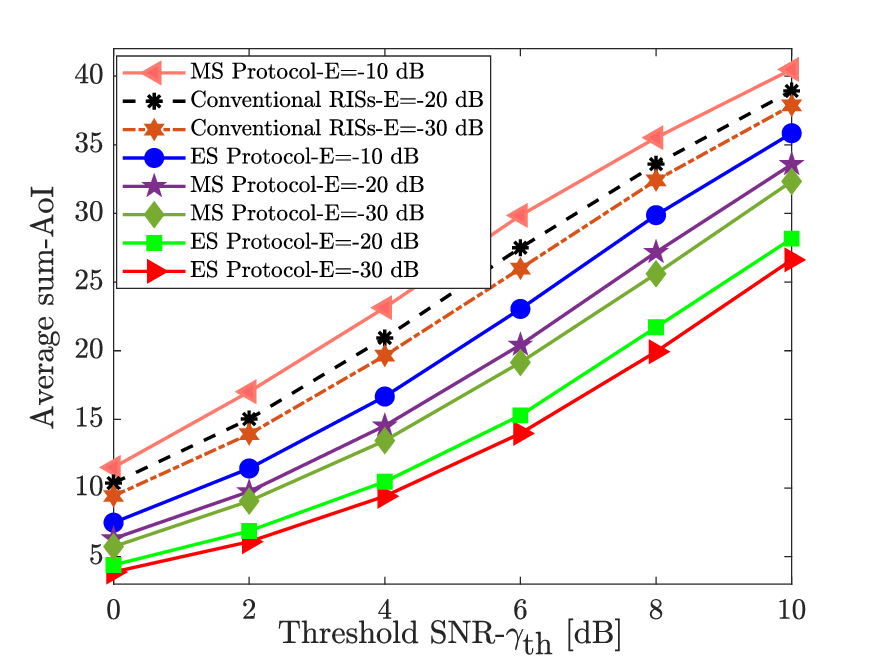}
    \caption{Average sum-AoI versus threshold SNR.}
    \label{fig:my_label2}
    \vspace{-.3 cm}
\end{figure}

Fig. \ref{fig:my_label2} depicts average sum-AoI versus SNR threshold $\gamma_{th}$ for $P_o = 3$, $M = 32$ and $N_t = 4$. Average sum-AoI increases with $\gamma_{th}$ because, with greater demands for communication quality, successful transmission is more challenging to achieve. Besides, the average sum-AoI increases by increasing the energy harvesting threshold from $E = -30$  to $E = -10$  [dB], since power allocation and beamforming must be adjusted to meet the requirements of energy users. So,  there is a compromise between the average sum-AoI and the harvested energy of energy users.
Additionally, compared to MS protocol, ES protocol shows better performance with the same parameters. Also, ES and MS based STAR-RIS always outperforms the conventional RIS structure.

\begin{figure}[t!]
    \centering
    \includegraphics[width=0.45\textwidth]{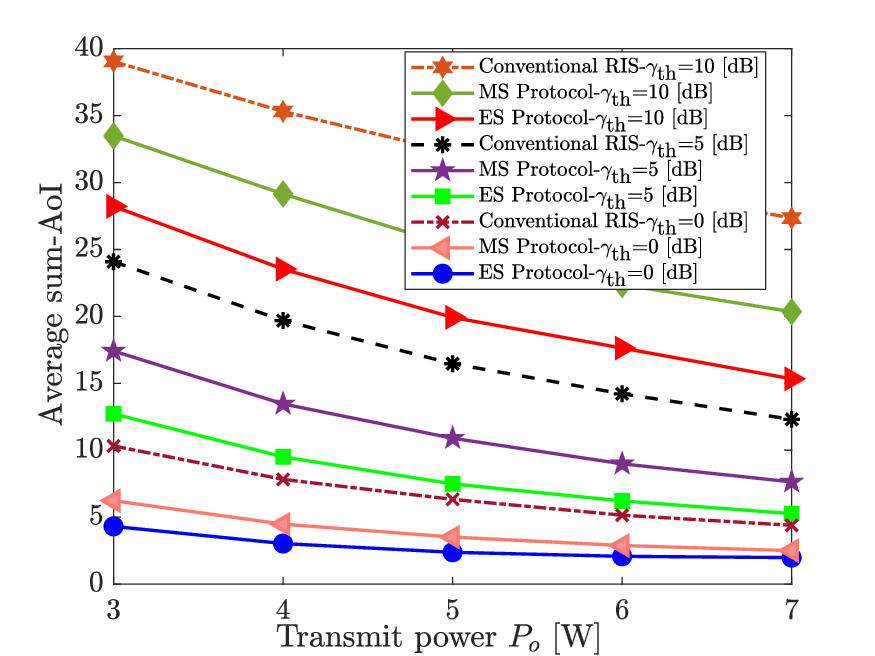}
    \caption{Average sum-AoI versus BS transmit power.}
    \label{fig:my_label3}
\vspace{-.3 cm}
\end{figure}

In Fig. \ref{fig:my_label3}, average sum-AoI versus transmit power $P_o$ is plotted for $E = -20$ [dB], $M = 32$ and $N_t = 4$. By increasing $P_{o}$, SNR is improved and due to higher chance of successful delivery, average sum-AoI is decreased. \textcolor{black}{Also, the difference between the performance of ES and MS protocols, and conventional RIS becomes more significant as $\gamma_{th}$ increases.}

\begin{figure}[t!]
    \centering
    \includegraphics[width=0.45\textwidth]{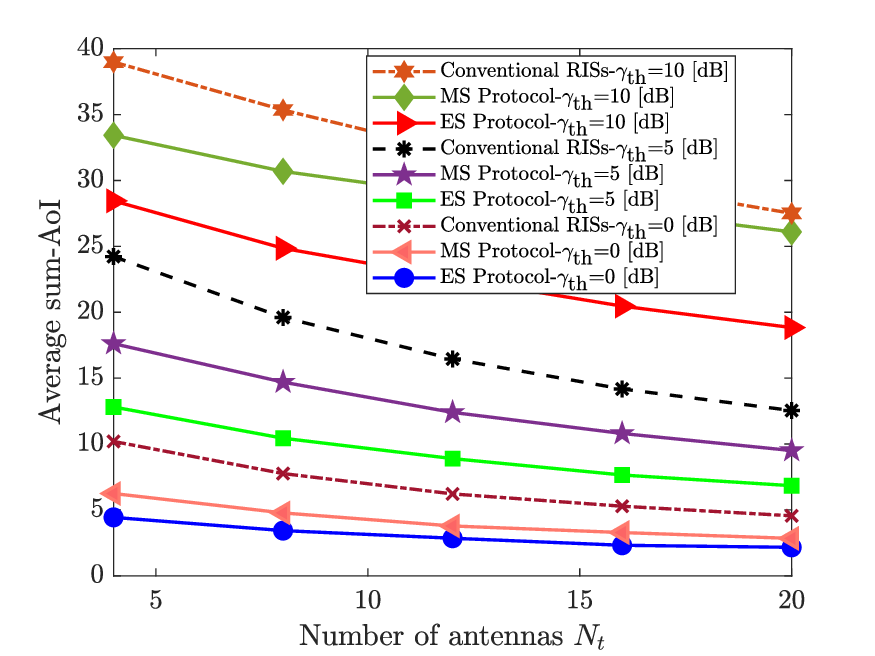}
    \caption{Average sum-AoI versus the number of antennas.}
    \label{fig:my_label4}
    \vspace{-.3 cm}
\end{figure}

\begin{figure}[t]
    \centering
    \includegraphics[width=0.45\textwidth]{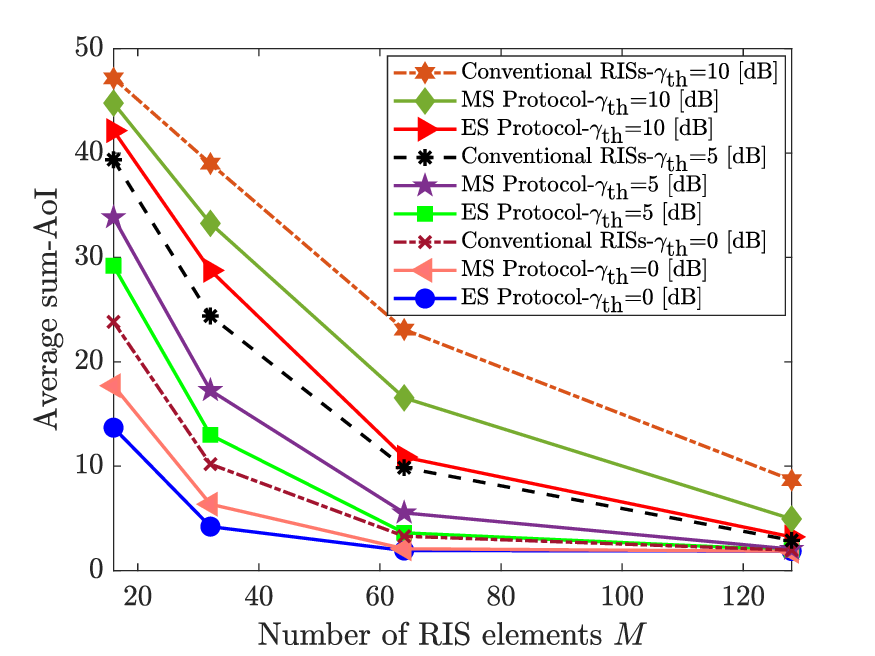}
    \caption{\textcolor{black}{Average sum-AoI versus the size of STAR-RIS.}}
    \label{fig:my_label5}
\end{figure}

The average sum-AoI versus number of antennas is illustrated in Fig. \ref{fig:my_label4}. It demonstrates that increasing the number of antennas $N_t$ from 4 to 20 at the BS improves the average sum-AoI, with $P_o = 3$, $E = -20$ [dB], and $M = 32$. Similarly, in Fig. \ref{fig:my_label5}, the average sum-AoI is decreased dramatically by increasing the number of STAR-RIS elements $M$ from 16 to 128, with $P_o = 3$, $E = -20$ [dB], and $N_t = 4$. \textcolor{black}{Therefore, based on the energy harvesting and data freshness requirements, one can select number of antennas at the BS and configure the STAR-RIS size and policy.}

\section{Conclusions}
We explored average sum-AoI optimization in SWIPT networks with the assistance of STAR-RIS. An AO algorithm was proposed to cope with the non-convexity of the optimization problem. STAR-RIS TaRCS are optimized by problem transformation, and beamforming vectors optimization at the BS is performed by SCA algorithm. Numerical results highlighted that the performance of the proposed algorithm for ES and MS protocols is more beneficial than that of the conventional RIS, and the ES protocol outperforms the MS one. Also, numerical results show a trade-off between the average sum-AoI and the harvested energy constraint. Increasing the number of antennas at the BS, 
 and especially the number of STAR-RIS elements can improve performance of the system.


\bibliographystyle{IEEEtran}
\bibliography{AoI_STAR-RIS}

\end{document}